# ENERGY CONSUMPTION MODEL IN AD HOC MOBILE NETWORK

Maher HENI[1], Ammar BOUALLEGUE[2] and Ridha BOUALLEGUE[3]

[1,3]Innov'Com Laboratory, Sup'Com, University of Carthage, Tunis, Tunisia
henimaher@gmail.com
ridha.bouallegue@gmail.com

[1,2]National School of Engineers of Tunis, University of Tunis El-Manar, Tunisia
ammar.bouallegue@enit.rnu.tn

## ABSTRACT

*The aim of this work is to model the nodes battery discharge in wireless ad hoc networks. Many work focus on the energy consumption in such networks. Even, the research in the highest layers of the ISO model, takes into account the energy consumption and efficiency. Indeed, the nodes that form such network are mobiles, so no instant recharge of battery. Also with special type of ad hoc networks are wireless sensors networks using non-rechargeable batteries. All nodes with an exhausted battery are considered death and left the network. To consider the energy consumption, in this work we model using a Markov chain, the discharge of the battery considering of instant activation and deactivation distribution function of these nodes.*

## 1. INTRODUCTION

Wireless ad hoc networks open various research perspectives since the posed constraints. Various proposed protocols in these complex networks that aim to reduce energy consumption or achieve self-configuration and self-organization. In addition messages sent must reach in real time accordingly to the applications nature of these networks.

Nodes are generally composed by embedded devices, with limited resources, autonomous, able to process information and transmit using radio waves, to another entity (sensors, processing unit, PDA...) over a limited distance. Ad hoc networks are generally composed by a large number of mobiles nodes, forming a large network without an established infrastructure. Therefore, the nodes act as transmitter, receiver, and data router at the same time.

Communications networks are used for various applications. Designing a network, it is necessary to evaluate its performances in order to correctly dimension its operations and capacity. Indeed, the evolution of these performances compares different network topologies according to their uses. It also allows the estimation of network behaviour in many situations, and in particular to estimate the operational problems and solve them before implementation. To evaluate the performances, then, we can use an analogy with a similar already existing network, use a network simulator or through analytical methods of verification. The networks simulators represent the advantage of uselessness of having existing networks. Another method of performances evolution of communications networks is the analytical analyses and formal verification. There are several evaluations methods using mathematical tools such as Petri nets [1], [2], [3], [4] or Markov chains [5], [6], [7]. The two methods present many limits with the modern complex system. Currently we would perform a model combining probabilistic and temporal aspect, to model a probabilistic systems variable in time. These two models (Petri and Markov chains) cannot provide a modelling formalism that satisfies these automata. To resolve these problems the researcher, analyses this complex system with model checking [8], [9]. This is a group of techniques for automatic verification [10] of dynamic system, since the varying of the system behaviour in time. Then it is important to express the timing and the probability in the same time.

To prefer the energy efficiency in mobile ad hoc networks, it is necessary to model the node energy consumption. In this work, our aim is to model the batteries power discharge, in ad hoc wireless networks. Our proposition is to use the IPP based ON /OFF model (Interrupted Poisson Process).This model is much used in various types of network to traffic modelling [11-18], even the use of such model in IEEE802.11 [19], have the aim to express the traffic control and not the energetic behaviour.

This paper is organized as follows: In Section 2, we present our motivation. We discuss the related work in section 3. The section 4 presents the stochastic modeling technics. In section 5 we present our proposition to model battery with ON/OFF model, and then we calculate the distribution probability function and the battery charge average. We conclude this paper and present the future work in section 6.

## 2. MOTIVATION:

A Mobile Ad-hoc Network (MANETs) [1] is a collection of nodes or terminals that communicate together by forming a multi-hop radio network. The nodes can move and their network topology may be dynamic.

In such network, there is no centralized administration. Each node can join the network or it leave at any time. Generally to ensure the mobility of nodes, we cannot recharge the battery. A node with an exhausted battery is considered dead and not usable.

In recent work [20] we presented an efficient routing method using energy efficiency parameters. With the exchange of considerable number of control messages to establishment of routes, or a data messages, this exhaust resources and aggravates performances. One of major causes of exchange of these message is the lost of paths, result of the exhausting of a node battery. To overcome this problem, we suppose that all nodes composed the networks have information about the energy stored in the batteries of its neighborhood. It avoid the routing using a node that it battery will be exhausted.

Precisely, a node uses at the same time the shortest selected path using Dijekstra [21] and also the residual battery energy in message routing. We therefore, proposed a new mechanism to send HELLO message with a reactive routing protocol (Ad hoc On-demand Distance Vector AODV [22], [23]) in calculated instant proportionally to the battery residual energy. Our proposition does not change the message fields, just with the sending instant we can extract the energy information. Reached the destination, calling the inverse function, it updates the neighbourhood energy base.

Through this information, a routing protocol avoids the use of a shortest path that is composed by a node with an exhausted battery.

## 3. RELATED WORK:

Given the growing importance of energy consumption in wireless networks, over the last few years, a number of works have proposed ways to improve energy efficiency and have modelled the terminals batteries in a wireless and wired network. So, in this context we present an overview on the batteries modelling, on the first hand, and the probabilistic model applied in wireless and wired networks.

In [35] the authors present the battery lifetime modelling using three different methods, the three models are used to predict the acid battery lifetime and look for means to extend it using numerical results.

In [24] and in their thesis report [25] M. R. Jongerden and B.R. Haverkort studied the mobile devices battery. Starting with a presentation of the chemical and electronic design, the authors express values of the charge and discharge procedures. Through an analytical study, they present in a second part the lifetime prediction of a battery using the Rakhmatov Vrudhila model [26]. This latter presents the diffusion process of the active material in the battery. In the third part of the former, the authors present the Kinetic Battery Model [27], [28], [29] it called such name because it uses a chemical kinetics process as its basis. The final part of [24] present the stochastic model using the Chiasserini and Rao [30], [31], [32], [33], that use the discrete time Markov chain to model battery. The different states used in this model represent the numbers of charge units. In the last section, authors use the stochastic modified KiBaM [34] using discrete time transient Markov process.

Our work is similar to this model. The difference is that our model does not present the N states as charge unit, but we use the two states of the node ON/OFF.

## 4. STOCHASTIC MODELLING:

The Markov chains [5], [6], [7] form a mathematical formalism to analyze a stochastic phenomenon. A model is based on states and transitions (transition rates or transition probability). Its simple structure allows modelling of a large class of systems. The Markov chains are well known tools which allow many theoretical results finely analyze the behaviour of the modelled system. The Markov chains are one of the most important tools for analysis of random processes in the area. As part of a modelling and performance evaluation, Markov chains have a simplicity and efficiency essential.

There are two major types of stochastic Markov chains with discrete time, in which time is considered discrete, consisting of a finite number of independent state, and continuous-time chain, formed by infinite state dependent and related. The states are the representation of random processes in time. Assume that the system transits from state i to state j with probability $P_{ij}$ that depends only on states i and j. Such a system is called without memory or Markovian, which means that the probability $P_{ij}$ does not depend on previous states i in which the system is passed over in its history.

To adapt a wide range of systems, we can use a Poisson process modulated by a Markov chain. The states are based on the intensity ($\lambda$) of a Markov chain. These processes are grouped under the acronym MMPP (Markov Modulated Poisson Process) [36]. An M-MMPP is an MMPP with M states.

The marginal distribution of MMPP is a combination of Poisson distributions. And its covariance is a weighted sum of exponentials. The model parameters are more important since we use with the transition matrix of the Markov chain, a new diagonal matrix denoted by $\Lambda$, containing the intensities associated with different states is necessary.

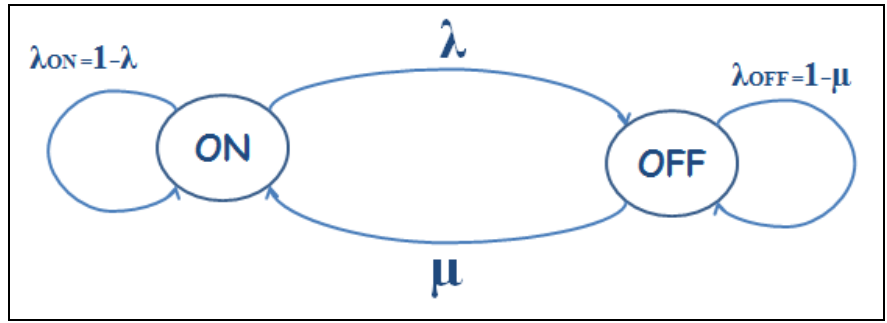

Figure 1.ON/OFF model: Markov chain with two states

ON-OFF model is a special case of MMPP and IPP (Interrupted Poisson Process). The Markov chain modulating the Poisson process has two states: ON and OFF, which correspond to two values of the intensity of a Poisson process (known as ON-OFF model because often in the off state intensity is zero). In addition to the 2x2 matrix of transition, this model requires two new parameters:

λon and λoff, which correspond to the respective intensities of the Poisson process in the ON and OFF states (see Figure.1). This model thus has four parameters (λon = (1-λ), λoff = (1-µ), λ, µ where λ and µ present respectively the transition probability from ON to OFF and from OFF to ON).

ON-OFF models are used to model the alternation of two phases or modes of communication with different characteristics. This allows reproducing some intermittency, which is a very important characteristic of traffic because it has a significant impact on performance.

In our modelling we will use this model to modulate the alternation between active and inactive state of a network nodes.

## 5. BATTERY ON/OFF MODELLING

### 5.1. Discharge Function

As mentioned in Section 2, we used the prediction of the battery usage of nodes that compose ad hoc networks in order to calculate optimal routes using as metrics energy and hop count. The model is well detailed in a previous work [20].

To complete the model we started from continuous distribution of battery power level presented in figure .2 According to [35].

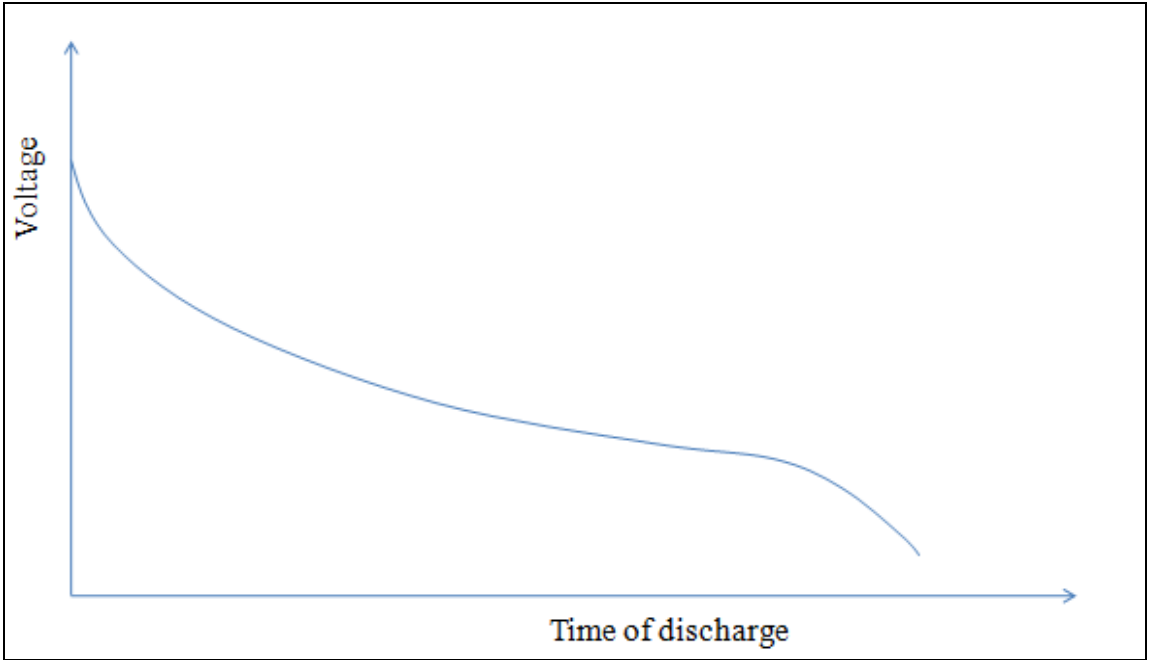

Figure 2.SOD: State of discharge

The figure .2 shows the state of discharge SOD presented by equation (1) that the demonstration found in [35]:

$$F = F_{init} + \int_{t=0}^{t} \frac{I_{SOD}}{C_N} d\tau \quad (1)$$

$$I_{SOD} = I - I_{GAS} \quad (1)$$

$$I_{GAS} = K \exp(C_U * V_N + C_U * T_N)$$

Where:
F : the state of discharge equation SOD.
$I_{SOD}$: discharge current
I: the charge and discharge battery current
$I_{GAS}$: the gassing current
K: constant depends on the battery parameters
$C_U$: voltage coefficient;
$V_N$: nominal battery voltage
$T_N$: nominal battery temperature

The both of calculation presented in [35], using the simplification, we choose a simple presentation of SOD equation:

$$I_{SOD} = K * \exp(\frac{-t}{\tau}) \quad (2)$$

## 5.2. Analytical model and Numerical results

In the real case, the curve shown in figure .1 is more accurate, since, there are times when the battery is inactive. By modeling the battery using the ON-OFF model, we will have a moment that no charge and no discharge of the battery as shown in figure .3.

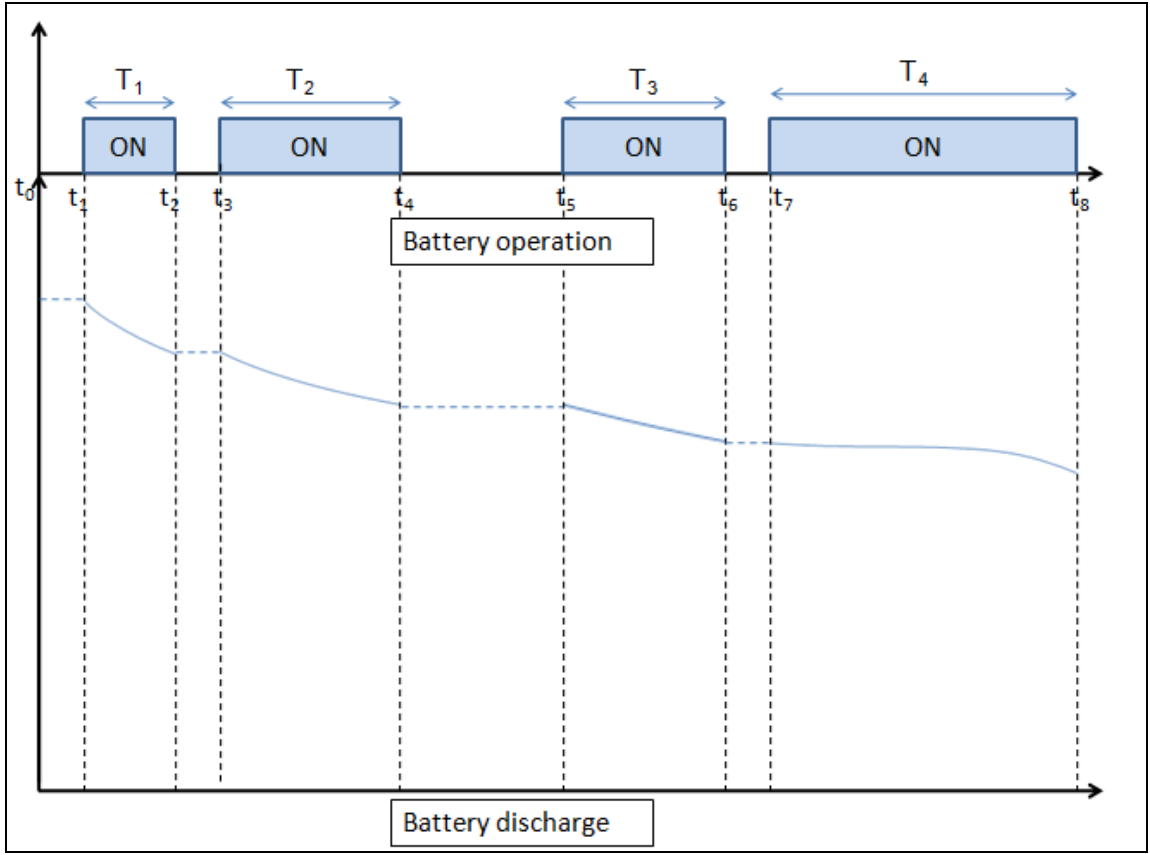

Figure 3. Battery discharge modulated ON/OFF

Our goal is to find the probability distribution function of time $t_i$ of battery operation.

Let $P(t, t+T) = e^{T*a_{ii}}$ the transition probability of the system from state i to state $j \neq i$, knowing that it passed a period T in the initial state.

Using the stationarity assumption "state transition probabilities are independent of the actual time at which transitions take place" [37, 38], P (t, t+T) is independent of t and $P(x_1, x_2) = P(0, x_2-x_1)$, then

$$P(t, t+T) = P(0, T) = e^{T*a_{ii}} \quad (3)$$

Using the ON/OFF model with the same parameters presented by figure .1, and we assume the system remains in the state ON during $T_i$ ($\forall\ i \in \{1,2,3...n\}$) and changes to the state off at time $t_i$ ( as shown in figure .3).
Let knowing that
$P_0$      the ON state probability
$P_1$      the OFF state probability
$T = T_1 + T_2 + T_3 + ... T_n \in [0, t]$

The transition probability is calculated as:
$$P = P_0(0,t_1) * P_1(t_1, t_1+T_1) * P_0(t_1+T_1, t_2) * ..... * P_1(t_n, t_n+T_n) * P_0(t_n+T_n, t)$$
$$= P_0(0,t_1) * P_1(0,T_1) * P_0(0, t_2-t_1-T_1) * ..... * P_1(0,T_n) * P_0(0, t-t_n-T_n)$$
$$= f_0(t_1) * f_1(T_1) * f_0(t_2-t_1-T_1) * ..... * f_1(T_n) * f_0(0, t-t_n-T_n)$$
$$= e^{-\mu t_1} * e^{-\lambda T_1} * e^{-\mu(t_2-t_1-T_1)} * ..... * e^{-\lambda T_n} * e^{-\mu(t-t_n-T_n)}$$
$$\cong \exp[-\mu*(t_1+t_2-t_1-T_1+t_3-t_2-T_2+...+t-t_n-T_n)] * \exp[-\lambda*(T_1+T_2+T_3+...+T_n)]$$
$$= e^{-\mu(t-T)} * e^{-\lambda T}$$

Then, using C as nearest constant we can proof equation (4)
$$P(T = \theta < t) = C * e^{-\mu t} * e^{-(\lambda-\mu)\theta} \quad (4)$$

To determinate C, we have integrated the probability on the entire interval
$$\int_0^t P(T = \theta < t) = 1 \Leftrightarrow C = \frac{\mu - \lambda}{e^{-\lambda t} - e^{-\mu t}} \quad (5)$$

Then the resulting probability is:
$$P(T = \theta < t) = \frac{(\mu - \lambda) e^{(\mu-\lambda)\theta}}{e^{(\mu-\lambda)t} - 1}$$

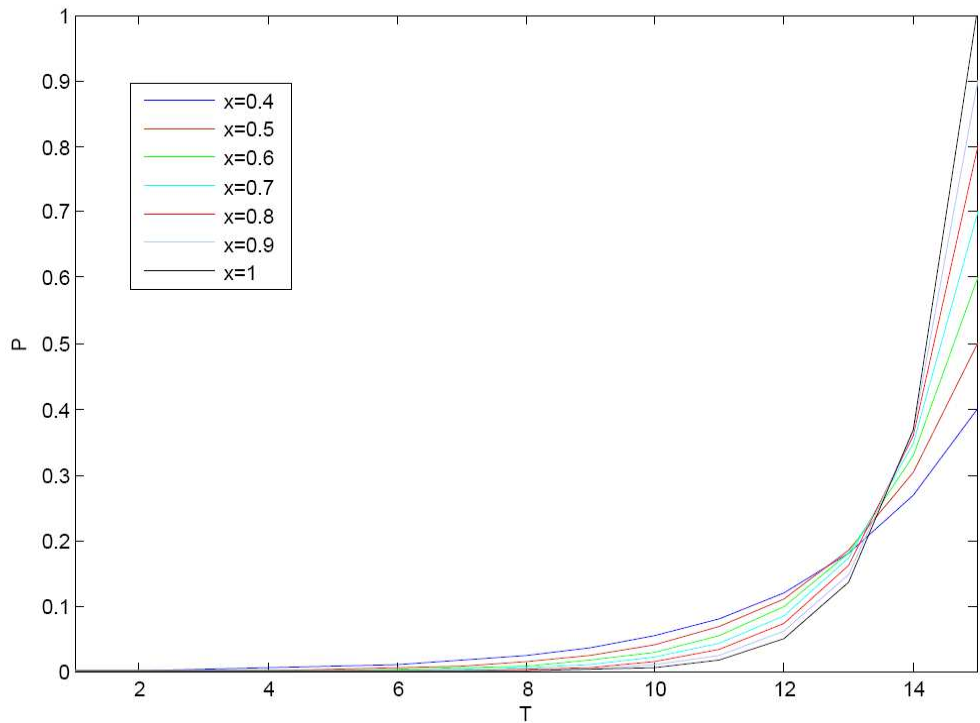

Figure 4. resultant probability

Figure .4 shows the different values of P, for x between 0.4 and 1, given that $x = \mu - \lambda$. For x = 1, the probability tends to 1 when T tends towards t. What it means, that the charge will run out if T = t, and if the battery pass the duration in the activate position.

The charge average at time T between 0 and t is presented by:

$$E[T] = \int_0^t \theta * \frac{(\mu-\lambda)e^{(\mu-\lambda)\theta}}{e^{(\mu-\lambda)t}-1} d\theta$$

$$E[T] = \frac{(\mu-\lambda)t+1}{(\mu-\lambda)} + \frac{t}{e^{t(\mu-\lambda)}-1} \qquad (6)$$

In figure .5 we plot the simulated average battery charge delay (presented by eq.6) for ON-OFF model. The bottom axis shows the transition rate between the two states ON and OFF (the $(\mu-\lambda)$ parameter).

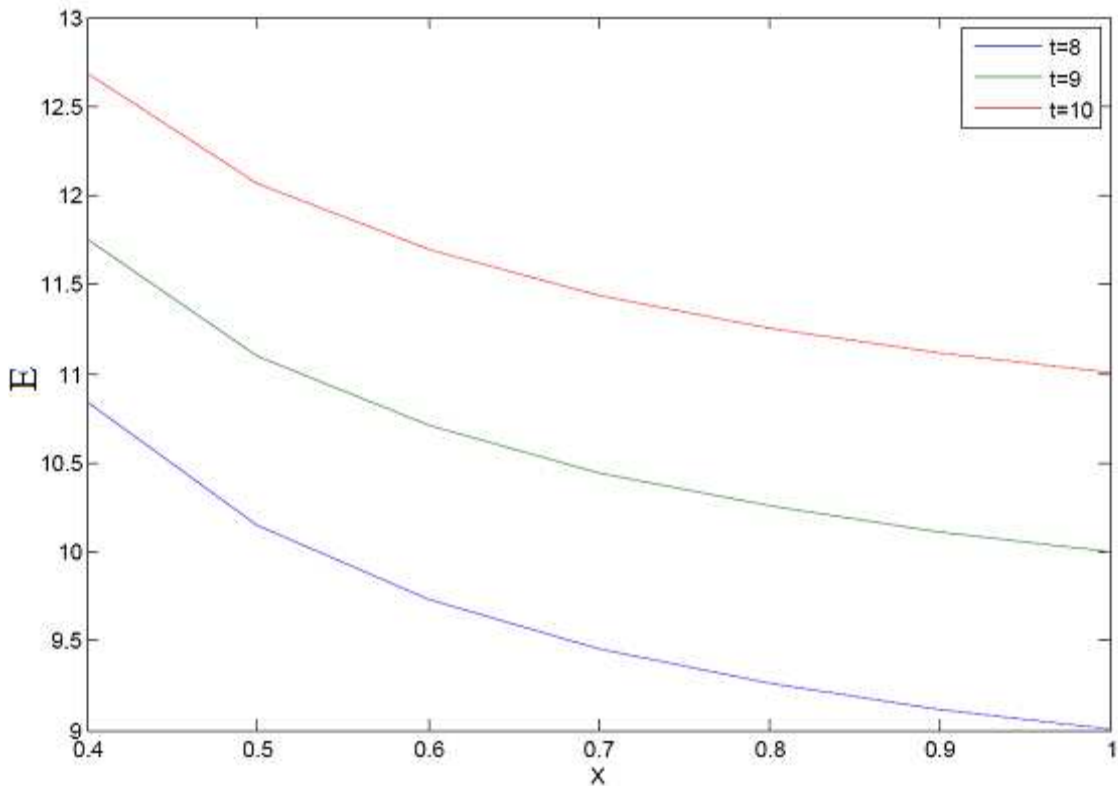

Figure 5. average of battery charge in t=T.

For the sake of simplicity, and without going far of reality, if we consider a battery with a standard activation duration D, in a normal conditions, the terminal using this battery will shut down after a duration D. However, if we alternate between start and shutdown of components, the nominal duration of battery will decrease. This can be translating by the both figure 4. and figure 5. The probability of operating of battery will increase if we decrease the number of alternation between activation and inactivation

($x = \mu - \lambda = 1$). Indeed, if this term decreases, T (the sum of activation periods) increase, and the battery duration will be near of D.

## 3. CONCLUSIONS

The power consumption of mobiles devices presents an important challenge in academic and industrial research. Much research has been done on decreasing the average power consumption of these devices. In this approach, the aim is to reduce the energy consumption by selective start or shutdown of components. However, in most of the research the battery is only considered as a limited-power source. Alteration between the active and inactive, and increased consumption due to state change are not taken into consideration.

In a routing algorithm, we proposed a routing of messages according to the level of the nodes batteries forming the path. The shortest and safest path according to battery charge nodes that form will be elected. To know the battery level of different nodes, the algorithm is to send HELLO messages in times depending on the level of the battery. The aim is to seek the function that reduces the maximum probability to have collision. The aim of this paper was to study the behavior of a nodes battery over a period T. Then, we chose as stochastic model Interrupt Poisson Processes IPP and especially the ON / OFF model. The calculation allows us to find a modeling of batteries. The variation and the calculation of averages favor the safety and correctness of our model. As future work, we will calculate the combined instant of sending messages and probability distributed function in order to find the right parameter and the optimal function that allows us to extract the battery level of a neighboring node, avoiding the collision.

**Authors**

**PhD. Maher HENI** was born in 1984 in KEF, Tunisia. He received the national engineering diploma in telecommunication from the National Engineering School of Tunis (ENIT) and the master diploma in Telecommunications from ENIT with collaboration of IEF institute in paris-sud university. Since 2010, he is a Ph.D. student in INNOV'COM (Innovation of COMmunicant and COoperative Mobiles Laboratory) in the Higher School of Communication (Sup'COM). His research interests are ad hoc networks and mobile communication systems.

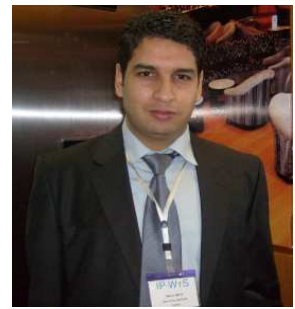

**Pr. Ridha BOUALLEGUE** was born in Tunis, Tunisia. He received the M.S degree in Telecommunications in 1990, the Ph.D. degree in telecommunications in 1994, and the HDR degree in Telecommunications in 2003, all from National School of engineering of Tunis (ENIT), Tunisia. Director and founder of National Engineering School of Sousse in 2005. Director of the School of Technology and Computer Science in 2010. Currently, Prof. Ridha Bouallegue is the director of Innovation of COMmunicant and COoperative Mobiles Laboratory, INNOV'COM Sup'COM, Higher School of Communication. His current research interests include mobile and cooperative communications, Access technique, intelligent signal processing, CDMA, MIMO, OFDM and UWB systems.

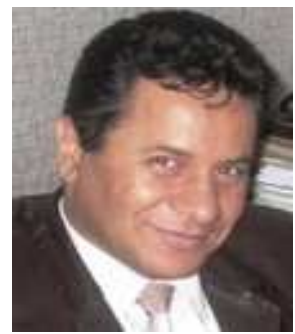

**Pr. Ammar BOUALLEGUE** was born in Kef, Tunisia, in 1945. He received the electrical engineering and engineer doctor degrees from ENSERG of Grenoble, France, in 1971 and 1976, respectively, and the Ph.D. degree from ENSEEIHT, INP of Toulouse, France, in 1984. In 1976, he joined the engineering school of Tunis (ENIT), Tunisia. From 1984 to 1992, he was the head of the Electrical Department, ENIT, and from 1993 to 1995, he was Director of Telecommunication for the High School of Tunis, Tunisia. He is the head of the Communication Systems (SysCom) Laboratory at the National Engineering School of Tunis. His research interests include passive and active microwave structures, signal coding theory, and digital communications.

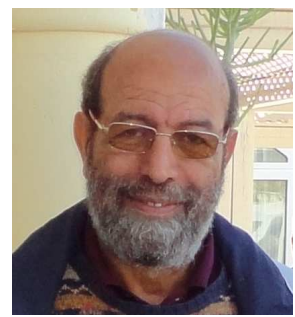